\documentclass[%
 reprint,
 amsmath,amssymb,
 aps
]{revtex4-2}

\nocite{*}
\usepackage{graphicx}
\usepackage{dcolumn}
\usepackage{bm}
\usepackage{float}
\usepackage{lineno}
\usepackage{listings}

\begin{document}

\preprint{APS/123-QED}

\title{Structural disorder on small scales softens hierarchical structures}
\author{Jonathan Michel}%
 

\author{Peter J. Yunker}
\email{peter.yunker@physics.gatech.edu}
\affiliation{%
School of Physics, Georgia Institute of Technology, 837 State Street, Atlanta, GA
30318, USA
}%

\date{\today}

\begin{abstract}
Hierarchically structured materials, which possess distinct features on different length scales, are ubiquitous in nature and engineering. In many cases, one structural level may be ordered while another structural level may be disordered. Here, we investigate the impact of structural disorder on the mechanical properties of hierarchical filamentous structures. Through simulations of networks with two hierarchical levels, we show that disorder does not change how stiffness scales with the mean coordination number - the average number of bonds per node - on large and small length scales. However, we find that network rigidity and stiffness depend strongly on the presence or absence of disorder on the small length scale, but not on the large length scale.  In fact, the amount of material necessary for a fully connected, network ordered on the small scale is insufficient to create even a marginally rigid network with small-scale disorder. We trace these phenomena back to a difference in the maximum mean coordination number on the small scale. While single length scale ordered and disordered networks have similar mean coordination numbers in the interior and on surfaces, we find that disorder strongly impacts the structure of surfaces, resulting in a larger fraction of surface nodes on the small scale. While this effect increases in strength as large scale bonds become narrower, it persists even for bonds that are wider than they are long (i.e., aspect ratios $ < 1$).
\end{abstract}

\maketitle



Hierarchically structured materials, in which small structures assemble into even larger building blocks, which then assemble into the final structure, are widespread in biological and engineered materials ~\cite{fratzl, lakes, wegst, rk}. They possess high strength to weight ratios ~\cite{wegst, meyers, harris, hier_expt}, can be both strong and tough ~\cite{launey, sen, ritchie} and exhibit a surprising robustness against fluctuations ~\cite{jmike}. In many cases, the smallest length scale is significantly smaller than the largest, such that the building blocks and their configuration on one scale may bear little relation to the next. In particular, many biological tissues are ordered on small length scales, but disordered on large length scales~\cite{wegst}. Ordered and disordered solids exhibit very different phenomenology, so it is unclear how a hierarchical structure with order on one length scale and disorder on another length scale may behave. Further, it is unclear why this motif -- ordered on short scales, disordered on long scales -- recurs so frequently~\cite{fratzl, wegst}.

Here, we seek to address this problem with a study of a model system in which slender,
elastic beams are assembled to form larger beams, which in turn are used to construct a larger
network. The use of elastic frames is well established as a tool for studying tissue mechanics
~\cite{broed2, cartilage},
and previous work has separately examined the mechanical properties of ordered hierarchical structures ~\cite{drk_auxetics,du,jmike} and disordered single length scale elastic networks ~\cite{wilhelm, lzhang} in detail. In the
present study, we synthesize these elements to understand the effect of geometrical disorder at
each scale. By geometrical disorder, we refer to a randomness in the placement of nodes in
networks, and accompanying disorder in the network topology and the orientation of bonds in the network.

At first glance, it may seem that some mechanical properties of hierarchical networks, such as the tensile stiffness, should not depend on the presence of order or disorder at all. After all, the stiffness of both ordered and disordered networks is controlled by the number of constraints present in the system above what is required for rigidity, i.e., the proximity to the isostatic point ~\cite{broed1}.
Through our simulations of two-length scale hierarchical networks of springs, we confirm that the presence of order or disorder on the large length scale has little impact on network stiffness.

However, we find that the presence of order or disorder on the small length scale has a major impact on both the onset of network rigidity and the maximum stiffness. To achieve rigidity, networks with disorder on the small scale require significantly more material than networks with order on the small scale. In fact, the amount of material necessary for a fully connected, network that is ordered on both length scales is insufficient to create even a marginally rigid network with small-scale disorder. This large material difference primarily results from the impact of disorder on surfaces. Disordered, hierarchical, filamentous networks have a systematically larger fraction of nodes on the surface of large scale bonds than either ordered hierarchical filamentous networks or single-scale disordered networks. As hierarchical networks necessarily have a substantial surface area on the smallest scales, we expect this effect to be present in many systems.

We study networks of slender beams, formed by first defining a large-scale
envelope, then filling this large-scale envelope with a small-scale network of
elastic filaments.
The structure of the large-scale envelope and small-scale network may be
defined in one of two ways: either a triangular lattice is formed, or a
Delaunay triangulation of a random point set is constructed. Random point
sets are Poisson disk packings, in which points are randomly distributed in space as densely as possible, while obeying the constraint that points be separated by a minimum distance, according to the algorithm described in~\cite{dunbar}. As a result, networks with order and disorder on the small scale are composed of springs with the same average size (i.e., $<1\%$ difference).

\begin{figure}
    \centering
    \includegraphics[scale=.3]{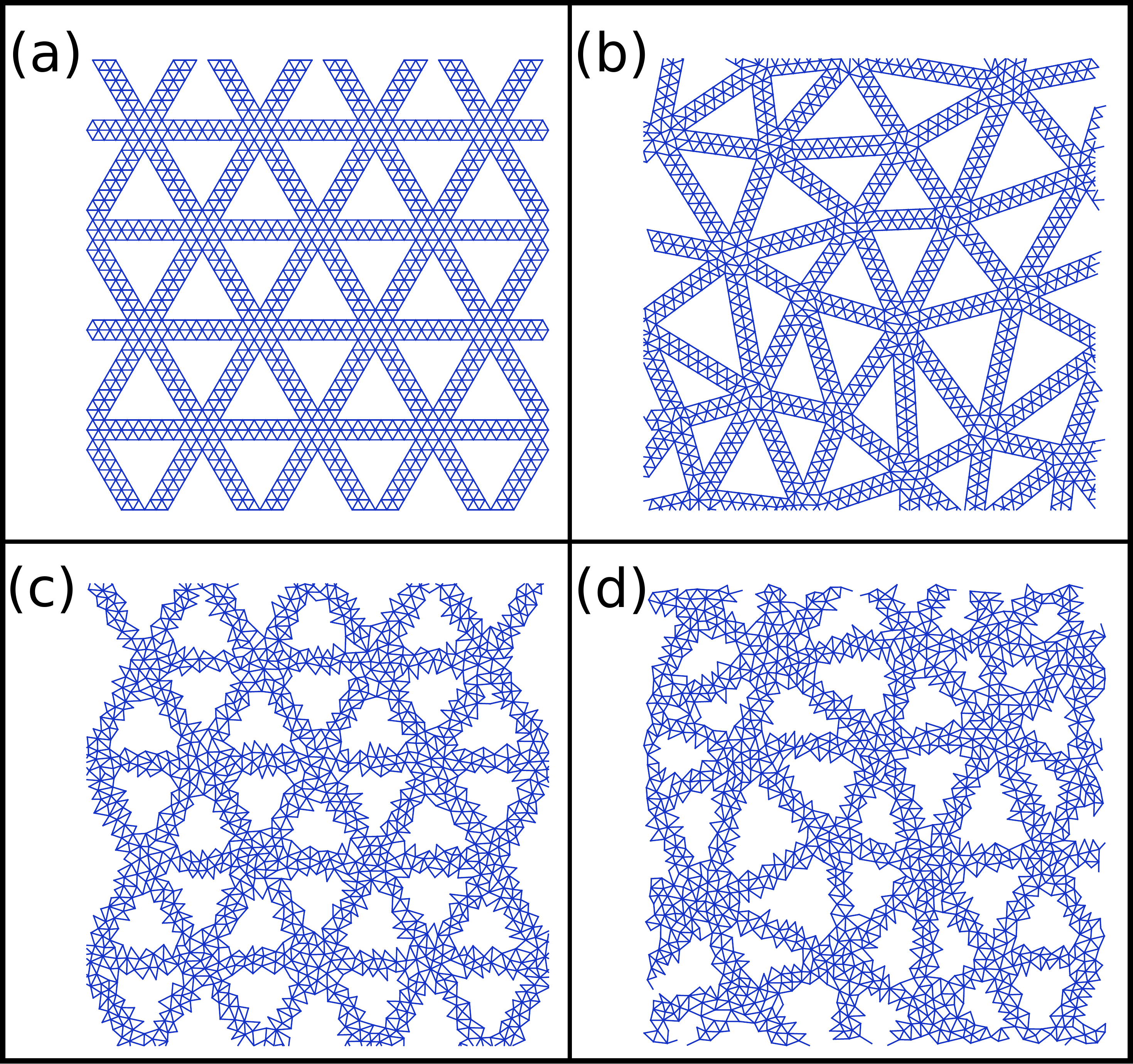}
    \caption{The four possible combinations of geometrical order and disorder: (a) A network with no geometrical disorder, (b) a network with large-scale geometrical disorder, (c) a network with small-scale geometrical disorder, and (d) a network with geometrical disorder on both scales.
    \label{netdiagram}}
    \label{fig:my_label}
\end{figure}

The small-scale may also be either crystalline, or formed from a Delaunay
triangulation of a Poisson packing. In the case of a crystalline small-scale structure and disordered large-scale structure, large-scale bonds are made of a regular small-scale triangular lattice, and stitched together at grain boundaries. Grain boundaries are created by identifying the areas in which multiple large-scale bonds overlap and triangulating these regions using the software Triangle~\cite{shewchuk}. Triangulations satisfy the constraint that internal angles of all triangles are at least $30^{\circ}$, and the area of each triangle does not exceed 1.5 times the area one of the equilateral triangular cells in the perfectly crystalline lattice. Schematics are provided in Fig.~\ref{netdiagram}.

Apart from varying the manner in which nodes are added to the network and connected to one 
another, we also consider random removal of bonds in the network at both the large and small
scales, such that a portion $p_l$ of large-scale bonds and $p_s$ of small-scale bonds are
retained. As discussed in~\cite{jmike}, we remove small-scale bonds in such a way that no
large-scale bond is cleaved, and that each pair of adjacent large-scale bonds is joined by
at least one small-scale bond. Removal of large-scale bonds is done in a manner that leaves
all large-scale vertices connected to the network. We expect that the decision to preserve a
single connected component will not greatly affect results, as we retain at least three fifths of
small-scale and large-scale bonds, and these fractions are well above the random percolation
threshold for triangular lattices of $\frac{1}{3}$.

Networks are modeled as ball-and-spring assemblies, and subjected to a
uniaxial tensile strain. Each small-scale bond is modeled as a fiber with
stretching modulus $\mu$, which, given a displacement field $\vec{u}$, has a stretching energy

\begin{equation}
E_{stretch} = \frac{\mu}{2} \int_0^l  \left|\frac{\partial \vec{u}}{\partial s} \right|^2 ds
\end{equation}

\noindent where $s$ is arc length and $l$ is the overall length of the fiber. Each fiber then has an effective spring stiffness $\frac{\mu}{l}$.

We simulated uniaxial tensile strains, in which the top nodes of a network
were displaced by varying amounts, and the vertical coordinates of top and bottom nodes were fixed, while all other degrees of freedom were
allowed to relax. Relaxation was carried out using the FIRE algorithm~\cite{fire}, and
was halted when the RMS of the residual forces on all degrees of freedom in
the network descended below $10^{-11}$, in units of fiber stretching modulus. To determine stiffness, we calculate the total strain energy, $E$, and fit a
power law of the form

\begin{equation}
\label{power_law}
E(s) = a \cdot \Delta y^b ,
\end{equation}

\noindent where $\Delta y$ denotes the displacement of the top nodes. Stiffness is then calculated as:

\begin{equation}
K = \frac{d^2 E}{d \Delta y^2}\Big|_{\Delta y = L / 1000} = a b (b - 1) \Delta y^{b-2} ,
\end{equation}
where $L$ denotes the overall length of the network. When fits of
~\eqref{power_law} yielded an $r$ value of less than 0.9, residual strain
energies after relaxation were exceedingly low - generally 8 orders of
magnitude or more below strain energies of fully connected networks - and
we regarded such networks as having zero stiffness.


For each combination of order and  disorder, we varied both large and small-scale bond portions from $0.6$ to $1$, to determine the role played by
large and small-scale structure in each case (Fig. 2).
Data once again are well-captured by a scaling model presented in
previous work ~\cite{jmike}, according to which the stiffness K is given by:
\begin{equation}
K = k \frac{(p_l - p_{l,c})(p_s - p_{s,c})}{(1-p_{l,c})(1-p_{s,c})}
\end{equation}
where $K$ is the tensile stiffness of a network, $k$ is the tensile stiffness
of a fully connected network, $p_l$ and $p_s$ denote the large and small-scale
bond portions, respectively, and $p_{l,c}$ and $p_{s,c}$ denote the critical
large and small-scale bond portions, respectively. In each case, we find that eq. 4 produces a good fit ($r^2 \approx .99$)

However, we find that the critical bond portions and maximum attainable stiffness depend strongly on the presence of order or disorder on the small length scale (see Table 1). Networks that are
geometrically disordered on the small scale first become rigid at a much higher small-scale bond portion. As a result, networks with small scale disorder require significantly more material than networks with small scale order to obtain the same stiffness. This effect is especially clear when considering the rigidity of networks fully connected on the large-scale (i.e., $p_l=1$) as the small-scale bond density increases, i.e., as more material is added (Fig. 3a). The attendant increase in density of small-scale bonds leads to a transition from zero to marginal stiffness at four distinct thresholds, with an earlier onset of stiffness in the presence of small-scale crystalline order. Notably, at a small-scale bond density sufficient for a network ordered on each scale to be completely connected, the network disordered on each scale is completely floppy. Beyond the transition region, stiffness increases approximately linearly with bond density, and at about the same rate for all four cases. However, the maximum attainable stiffness (i.e., the stiffness at full connectivity) varies across all four cases, and is significantly lower for networks with small scale disorder.
\setlength{\extrarowheight}{1pt}
\begin{center}
\begin{table}
\begin{tabular}{|l|c|c|c|c|}
     \hline
     Large / Small Scale & k & $p_{l}$ & $p_{s}$ & $R^2$\\
     \cline {1-5}
     Order/Order & .410 & .58 & .83 & .989\\
     \cline{1-5}
     Disorder/Order & .364 & .62 & .82 & .988\\
     \cline{1-5}
     Order/Disorder & .244 & .62 & .89 & .990\\
     \cline{1-5}
     Disorder/Disorder & .230 & .63 & .89 & .989\\
     \cline{1-5}
\end{tabular}
\caption{Maximum attainable stiffness, large and small-scale bond critical portions, and $R^2$           values for agreement between our scaling ansatz and simulation data are shown for each
     combination of large and small-scale structure. Crucially, structures that lack small-scale
     crystalline order require a far higher portion of small-scale bonds to be retained, and are
     significantly softer than their crystalline counterparts, even when fully connected.}
\end{table}
\end{center}



\begin{figure}
    \centering
    \includegraphics[width=.5\textwidth]{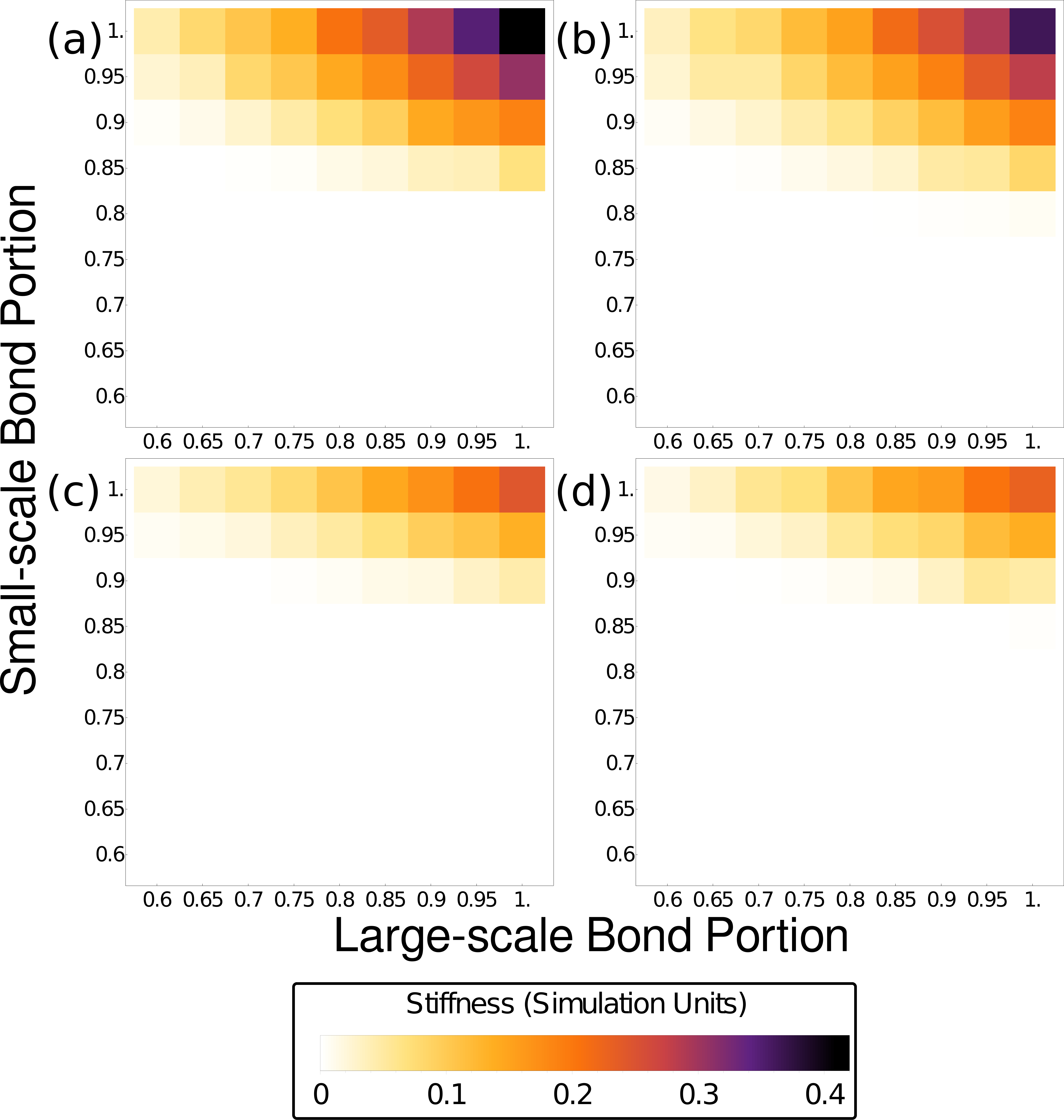}
    \caption{Stiffness vs. large and small-scale bond portion is shown for
    (a) no geometrical disorder, (b) geometrical disorder on
    the large scale, (c) geometrical disorder on the small scale, and
    (d) geometrical disorder on both scales. We note that maximum
    attainable stiffness values are much lower, and the minimum small-scale
    mean coordination number needed for marginal stiffness is much higher, when networks
    are disordered on the small-scale.}
\end{figure}

To understand why the onset of rigidity and maximum attainable stiffness both depend strongly on the presence of order or disorder on the smallest scale, we seek to capture the dependence of $K$ on the mean coordination number. We plot $K$ versus $(z_l - z_{l,c})(z_s - z_{s,c})$ where $z_l$ and $z_s$ denote the mean coordination number of large and small-scale bonds, respectively, $z_{l,c}$ and $z_{s,c}$ represent the mean large and small-scale coordination numbers necessary for marginal rigidity. Use of a scaling model of this sort for a lattice with a single characteristic length scale is well established ~\cite{ohern, goodrich}. Here we combine this approach with our previously presented model for hierarchical networks (described above and in~\cite{jmike}). We find that the product of excess coordination numbers on large and small scales captures simulation data for each class of network considered. Further, we find that the critical coordination numbers are $\sim$~3.5 on the large scale (likely due to an emergent bending rigidity ~\cite{das,mao,gab,jmike}), and $\sim$~4.0 on the small scale (as expected from Maxwell Counting) for all cases. Instead, it is the maximum mean coordination numbers (i.e., the mean coordination number of fully connected networks with $p_s=1$ and $p_l=1$) that vary.

\begin{center}
\begin{table}
\begin{tabular}{|l|l|l|l|}
     \cline{1-4}
     Large / Small Scale & $z_l$ & $z_s$ & $R^2$\\
     \cline {1-4}
     Order/Order & 3.44 & 4.09 &  .994\\
     Disorder/Order & 3.52 & 3.94 & .990\\
     Order/Disorder & 3.42 & 4.10 & .997\\
     Disorder/Disorder &3.44 & 4.01 & .995\\
     \cline{1-4}
\end{tabular}
\caption{Data are shown for the critical large-scale and small-scale mean coordination numbers for
each combination of order and disorder.}
\end{table}
\end{center}

To understand the variation in maximum mean coordination numbers, we considered single length scale ordered and disordered networks, with heights equal to their widths. The maximum mean coordination number for these non-hierarchical networks is 6, regardless of the presence or absence of disorder. This follows from the fact that a triangulation of $n$ points with $k$ points on
its convex hull has $3n - 3 - k$ edges ~\cite{compgeom}. As $n$ becomes large, the number of
edges per node will tend to 3, so that mean coordination number will approach 6. Further, crystalline
networks cleaved along boundaries aligned with one of their symmetry axes will have boundaries
with a mean coordination number of 4, and when random networks are cleaved along a random line,  we have found that nodes along this line will have a mean coordination number of 4.04 (see SI for more details on calculating the mean coordination number on surfaces). The difference in maximum mean coordination number is therefore not due to intrinsic differences between bulk properties of ordered and disordered networks.

\begin{figure}
    \label{stiffness_dens_z}
    \centering
    \includegraphics[width=.45\textwidth]{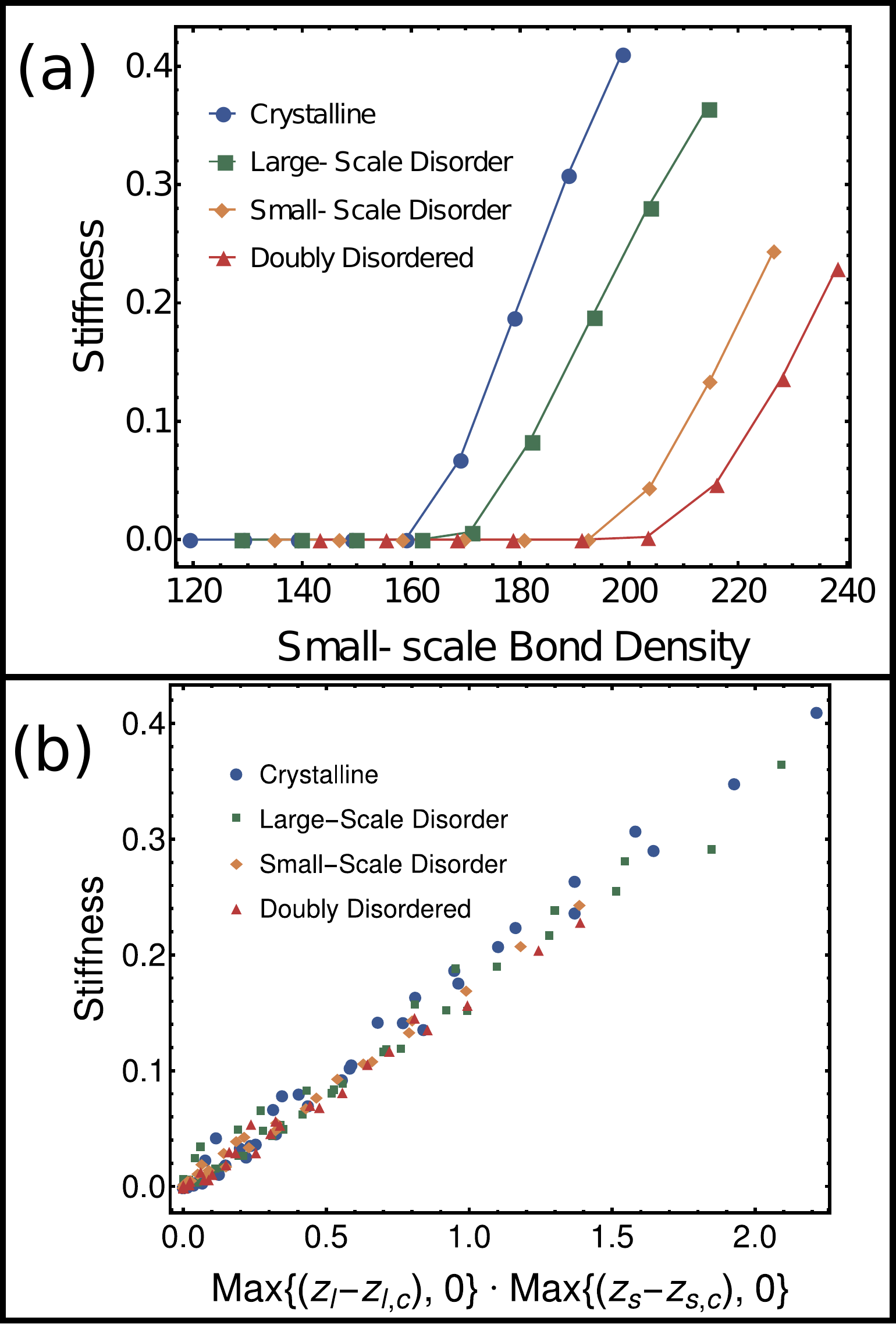}
    \caption{(a) A plot of network stiffness versus the density of bonds on the small-scale reveals that geometrical disorder delays the onset of rigidity with increasing
    numbers of bonds, and beyond this threshold, limits the additional benefit conferred by
    increasing the density of bonds. Notably, rigidity transition points to marginal stiffness are
    clustered according to small-scale structure. (b) Stiffness scales in a similar manner with
    respect to the product of excess large and small-scale mean coordination number for all four cases.}
\end{figure}

\begin{figure}[H]
    \centering
    \includegraphics[width=.4\textwidth]{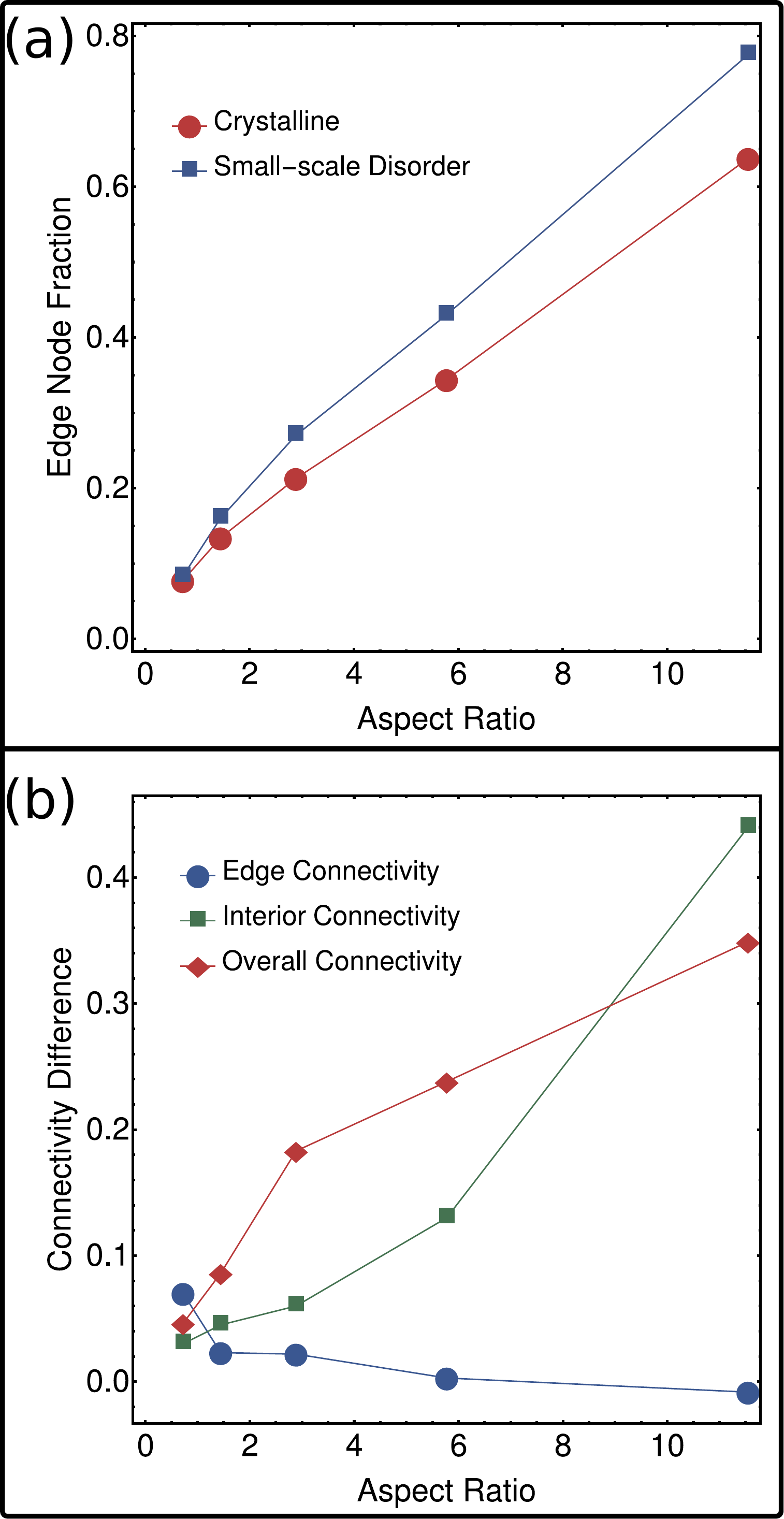}
    \caption{We compare fully connected networks with large-scale crystalline
    order and either small-scale crystalline order or small-scale disorder in
    terms of (a) the fraction of vertices that are on the perimeter of a
    large-scale bond, or (b). the difference in mean surface coordination number,
    mean interior coordination number, and total mean coordination number, for varying large-scale
    bond aspect ratio $\alpha$. While ordered and disordered slender large-scale bonds exhibit a marked disparity in edge fraction, as well as in
    interior and overall mean coordination number, edge node fraction and mean coordination number maintain small, but non-zero differences for wide large-scale bonds. We find that this disparity makes an
    essential contribution to the marked decrease in stiffness resulting from
    the loss of crystalline order.}
    \label{fig:boundary_study}
\end{figure}

In fact, the difference in mechanics across these scenarios primarily arises due to variation in \textit{how many} nodes are on bond surfaces. Networks with small-scale disorder have a systematically larger fraction of nodes on the surfaces of bonds (Fig. 4A). We demonstrate this
by computing the fraction of small-scale vertices on the edges of large-scale bonds for networks
with large-scale bonds twenty times the length of small-scale bonds, for varying aspect ratio,
$\alpha$, where $\alpha$ is defined as the ratio of bond length to bound width. The disparity in
the fraction of nodes located at the edge persists even as bonds grow wider, though the magnitude
of the effect diminishes (Fig. 4A \& B). We find this disparity is the primary driver of the
difference in overall mean coordination number between filamentous networks with small scale order and
disorder.

Note that, while the surface mean coordination number is quite similar in small scale ordered and disordered networks, the interior mean coordination number varies for networks with large aspect ratios (Fig. \ref{fig:boundary_study} B). However, the fraction of interior nodes declines precipitously as the difference in interior mean coordination number grows. Thus, while the difference in interior mean coordination number does affect the difference in overall mean coordination number, it has a smaller effect than that of the difference in the fraction of surface nodes.



To check if these results arise due to the manner in which we constructed large scale bonds with small-scale disorder, we repeated our measurements with a different construction technique. We still construct networks via a Delaunay triangulation of a Poisson packing, but we now insist on keeping the node density equal to that of a lattice with small-scale order (rather than keeping bonds the same mean length). As a result, small scale bonds in networks with small-scale disorder have a mean length about 1.03 times that of bonds in networks with small-scale order. While our quantitative results change, this modification does not impact the qualitative results (see SI for more details about this alternative construction). While we expect that a network could be constructed by hand to possess geometric disorder and a similar portion of surface nodes, doing so would be non-trivial. For example, one could take a network that is ordered on the small-scale, preserve its bond topology but randomly shift the positions of each node. However, through the point at which bonds begin to cross (which is non-physical), a non-zero degree of orientational order persists (see SI for more detail).


The work presented here suggests that, though the large and small-scale coordination number
contribute to the overall mechanical properties of the material in mathematically analogous ways, well-controlled small-scale assembly is most important for determining the onset of rigidity and the maximum stiffness. This phenomenon is manifest in hierarchical structures; ordered and disordered single-length structures studied here obtain rigidity with a similar amount of material, and exhibit similar maximum stiffnesses. When large-scale bonds are made of a disordered solid, we find that a profusion of under-coordinated edge bonds is likely to compromise the rigidity and attainable stiffness of the material. Moreover, networks with crystalline structure on the small scale ensure greater return on investment of material as the number of connections is increased. Thus, the results presented here may provide a guide for designing and constructing hierarchical materials.

Further, these results generalize the previous finding that ordered hierarchical networks are robust against random assembly errors. Here, it is demonstrated that a simple extension of a class of mean field model developed for networks with a single characteristic scale captures the data presented here, giving further credence to the idea that a simple scaling law can account for many mechanical properties of hierarchical systems~\cite{jmike}. As a result, the same emergent robustness is present in hierarchical networks that lack orientational and positional ordering on small or large-scales.

In the context of biological tissues, this work suggests that robust assembly of the ordered, basic building blocks is most important, and that later stages of assembly may be more forgiving of imprecision. Ultimately, constructing a tissue from ordered building blocks requires less material than constructing a tissue from disordered building blocks. While order on the smallest scales may emerge as a simple consequence of molecular self-assembly, the economy of material could itself serve as a strong incentive for evolving a modular assembly process with a particularly robust, ordered initial stage, as it allows resources and energy to be allocated to other tasks.

\bibliographystyle{apsrev4-1}
\bibliography{apssamp}

\end{document}